\documentclass[
  aps,
  pre,
  nofootinbib,
  longbibliography
]{revtex4-2}

\usepackage{multirow}
\usepackage{lipsum}
\usepackage{amsfonts}
\usepackage{graphicx}
\usepackage{epstopdf}
\usepackage{algorithmic}
\ifpdf
\DeclareGraphicsExtensions{.eps,.pdf,.png,.jpg}
\else
\DeclareGraphicsExtensions{.eps}
\fi

\usepackage{subfigure}
\usepackage{graphicx}
\usepackage{amsmath,amsfonts,amssymb}
\PassOptionsToPackage{dvipsnames}{xcolor}
\usepackage[dvipsnames]{xcolor}
\usepackage{comment}
\usepackage{tikz}

\usepackage{float}
\usepackage[inline]{trackchanges}
\addeditor{LK}

\begin{document}

\title{Topological Data Analysis combined with Machine Learning for Predicting Permeability of Porous Media}

%\author{E. Dagdelen$^{1}$, C.N. Lalu $^{2}$, A. Karlekar$^{1}$,  M. Arora $^{1}$, M. Illingworth$^{1}$, J. Jaquette $^{1}$, L.J. Cummings$^{1}$ and L. Kondic$^{1}$}
%\thanks{}
%\affiliation{$^{1}$ Department of Mathematical Sciences, New Jersey Institute of Technology, Newark, NJ\\
%$^{2}$ Department of Physics, New Jersey Institute of Technology, Newark, %NJ
%}

\author{E. Dagdelen$^{1}$, C.N. Lalu $^{2}$, A. Karlekar$^{1}$,  M. Arora $^{1}$, M. Illingworth$^{1}$, J. Jaquette $^{1}$, L.J. Cummings$^{1}$ and L. Kondic$^{1}$}
\thanks{First and second author contributed equally}
\email[Corresponding author:]{kondic@njit.edu}
\affiliation{$^{1}$ Department of Mathematical Sciences, New Jersey Institute of Technology, Newark, New Jersey 07102, USA\\
$^{2}$ Department of Physics, New Jersey Institute of Technology, Newark, New Jersey 07102, USA
}

\date{\today}

\begin{abstract}
Flow in porous media is challenging to address using standard analytical or numerical methods due to the geometrical complexity.  However, since synthetic representations of porous media are easy to produce and data from physical experiments are becoming more widely available, the problem is well-suited to studies that include machine learning (ML) techniques.  We discuss a number of features that can be extracted from such synthetic or experimental data, and their utility as input variables into a standard ML algorithm.  These features include structural measures describing the geometry of the porous media, topological measures describing the connectivity, and network measures obtained by modeling the porous media as simplified pore networks.  These features enable the prediction of the permeability of the considered (synthetic) porous materials using ML techniques that also leverage the separately computed exact permeability (ground truth).  Comparing results obtained using different input variables helps develop a better understanding of the utility of various measures for predicting permeability based on the porous media structure.  We show, in particular, that topological data analysis (TDA) provides a useful set of features that can be easily combined with ML to yield meaningful results.
\end{abstract}
\maketitle

\section{Introduction}

Machine Learning (ML) already plays an important role in materials science research, and their use is expected to expand significantly in the years to come.  The goal of the present work is to provide a simple description of selected approaches for analyzing porous media, specifically problems involving flow through porous media, which are important in a variety of applied fields. This system is a good candidate for ML-based approaches, since the exact (numerical) solutions are hard to compute due to the computational complexity; and the underlying data (digitized representations of porous media) can be easily generated, as in the present work, or obtained from experimental images. Extensive studies of such porous media systems, both with and without consideration of ML, have been carried out (see, e.g.,~\cite{Blunt2013PoreScaleModelling, robins2011theory, VanDerLinden2016, suzuki2021flow,elmorsy2021water,Zhang2024}); the distinguishing feature of the present work is that it (i) presents new results illustrating the value of topological data analysis (TDA), particularly those emerging from persistence homology~\cite{mischaikow} in the context of porous media flow and ML; (ii) presents the techniques used in an accessible manner, avoiding complexities and technical details, which, while possibly important for any particular application, may obscure the big picture; (iii) uses well-documented software packages and libraries that can be implemented with minimal preparation; and (iv) provides detailed, clear, and precise measures that quantify the utility of ML in the present context.  To illustrate one possible application of the present work, let us point out that a significant part of the techniques and some preliminary results emerged from a semester-long undergraduate (Capstone~\cite{web_capstone}) project in the Department of Mathematical Sciences at New Jersey Institute of Technology; the senior authors hope that the presented paper may be useful to other educators and researchers looking for a material that can be used to introduce students to the ML and TDA techniques, which they may find useful in their future professional lives, either in an academic or industrial environment.  

Permeability is a fundamental property of a porous material that quantifies the ease with which fluid can pass through it under pressure. It plays a crucial role in Darcy’s Law, which states that the volumetric flow rate of a fluid through a porous medium is directly proportional to the pressure gradient and the medium's permeability, and inversely proportional to the fluid's viscosity; see~\cite{probstein} for a concise review. Accurate estimation of permeability is essential in a wide range of real-world applications, including groundwater filtration, oil recovery, CO\textsubscript{2} sequestration, and industrial filtration processes~\cite{Korvin2024}. In these examples, the porous material is soil, rock, or some manufactured porous membrane. Porous media have been studied using a variety of modeling approaches to understand how fluids flow through the complex internal pore structures. We note that porous media flow is closely related to percolation theory; see~\cite{feder88,stauffer} for standard physics texts on this topic,~\cite{hunt2014} for discussion of aspects particularly relevant to porous media flow, or~\cite{bollobas2009} for a more mathematical perspective. Standard methods for modeling porous media flow include continuum and pore-scale models, which describe fluid transport using principles such as Darcy’s law and the Hagen–Poiseuille equation, and can be extended to capture additional phenomena that may be important in applications, such as pore blockage, adsorption, and the accumulation of deposits over time \cite{Berg2023PercolationMedia, Sparks2016FiltersHandbook}. Pore-network models provide a simplified representation of a porous medium as a network of interconnected pores of simple shape. Being more analytically tractable, such models can yield additional insight, allowing systematic exploration of how pore size, connectivity, and heterogeneity affect overall fluid transport through the medium~\cite{Cummings2026FiltrationNetworks,FattECharacteristicsIII, griffiths-jcis-2014, gu-jms-2022}.

Traditionally, permeability is measured through laboratory experiments (see~\cite{velasquez2024phase} for an example of a recent work in this direction) or by simulating fluid flow through domains obtained from digitized images of a porous medium, using partial differential equations (PDEs). Although accurate, such PDE-based simulations are computationally intensive, often requiring fine-grained meshes, high-resolution image data, and substantial processing time to converge, especially for large or structurally complex samples. These challenges have motivated growing interest in data-driven alternatives and advanced tools that can efficiently and accurately account for realistic pore structures~\cite{eage,graczyk,robins2011theory}. One approach proposed recently applies persistent homology (PH) (see~\cite{porter2023phystoday} for a recent, accessible review, and ~\cite{carlsson,edelsbrunner-harer} for more in-depth studies), a tool from topological data analysis (TDA), to study the structure of pore networks~\cite{Illingworth26,suzuki2021flow}. Such methods provide a potential foundation for data-driven approaches, including machine learning (ML), to efficiently predict permeability in complex porous materials.

In this work, we investigate whether an ML model trained on geometric, topological, and network-based descriptors of a digitized representation of a porous medium can accurately predict its permeability. We begin our workflow by generating 1000 synthetic porous structures using the Porous Microstructure Analysis (PuMA) software package \cite{puma}. For each of these structures, we compute permeability directly using PuMA's built-in capabilities. We then extract pore network representations from the structures using the SNOW2 algorithm implemented in PoreSpy \cite{porespy}. Moreover, for each structure and its associated pore network, we compute a comprehensive set of descriptors that capture geometric complexity, topological features, and network connectivity.

The combined structural, topological, and network descriptors are then used to train the ML model to predict permeability. Permeability prediction using ML can significantly reduce the computational cost and run time of traditional methods while retaining physical interpretability. The objectives of the study are 
(i) to assess the predictive power of these structural, topological, and network descriptors, and (ii) to evaluate the computational efficiency of ML predictions relative to direct numerical simulations. Here, accuracy is measured by comparing with permeabilities computed with PuMA, while efficiency is evaluated by total computation time. Whenever possible, we attempt to describe the techniques used and the results in a simple yet detailed manner, allowing for a reasonably straightforward understanding and reproducibility.

The rest of this paper is structured as follows. Section~\ref{sec:methods} discusses our methodology, starting with data generation in Section~\ref{sec:data}, followed by a description of how various topological, network, and structural features are extracted from the data in Section~\ref{sec:feature} (here, we also provide toy examples that illustrate the applied methods in a detailed but easy-to-understand manner).  We next provide an outline of how permeability is computated in Section~\ref{sec:permeability}, then a discussion of the ML neural network model implemented in Section \ref{sec:ML}. Section~\ref{sec:results} includes our main results; we complete this study with the Summary and Conclusions, in Section~\ref{sec:conclusions}.

\section{Methods}
\label{sec:methods}

In this section we first describe how our artificial porous media domains are constructed in Sec.~\ref{sec:data}, then discuss how the various features used to train the ML model are extracted in Sec.~\ref{sec:feature}. Section \ref{sec:permeability} describes how the ``ground truth'' permeability of the structures, with which we compare our ML results, is obtained. The details of how the ML model is built, trained, and evaluated are discussed in Section~\ref{sec:ML}.

\subsection{Data Generation}
\label{sec:data}

\begin{figure}[tbp]
    \centering
    \subfigure[]
        {\includegraphics[width=0.49\linewidth]{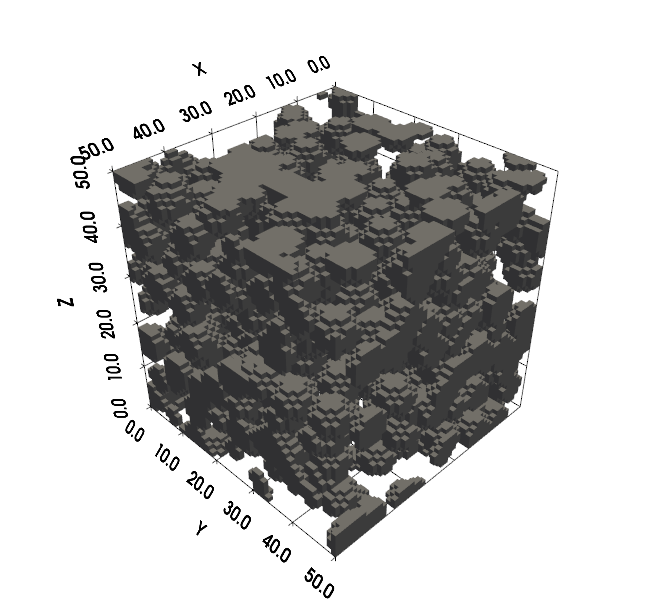}}
    \subfigure[]
        {\includegraphics[width=0.4\linewidth]{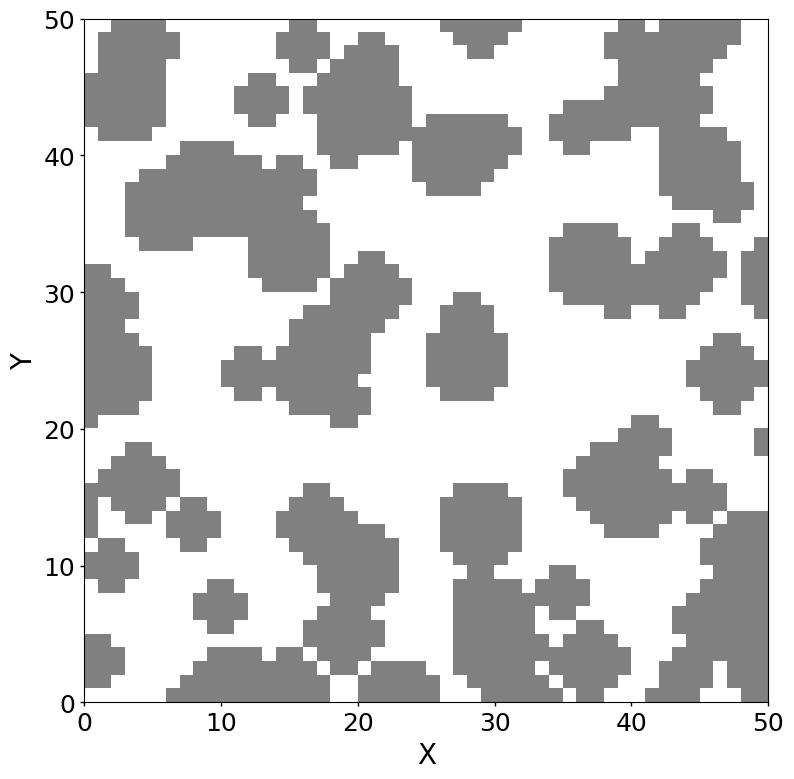}}
    \caption{
    (a) 3D visualization of a typical synthetic porous medium dataset of $50 \times 50 \times 50$ voxels generated using PuMA software; 
    (b) 2D cross-section of the same simulated 3D dataset, perpendicular to the $z$-axis. Gray regions correspond to solid material and white regions to void space.}
    \label{fig:puma_example}
\end{figure}

To facilitate ML-based permeability prediction, we use PuMA~\cite{puma} to construct 1000 synthetic porous structures composed of randomly-distributed solid spheres. The spheres represent solid obstacles, while the spaces between them serve as the void space through which fluid can flow \cite{roding2020}. PuMA allows the generation of overlapping spheres with user-defined parameters, enabling control over porosity and pore size in the synthetic structures.   In our work, the sphere diameters range from 5 to 15 voxels, and the target porosity is set to 0.5. Each structure is created by placing the spheres at random positions within a cubic domain of 50$\times$50$\times$50 voxels (computations on larger domains were explored, but for computational efficiency, we limit this study to the 50$\times$50$\times$50 domain).  Figure~\ref{fig:puma_example} shows the 3D visualization of a typical simulated synthetic porous medium dataset and a corresponding 2D cross section perpendicular to the $z$-axis, where gray regions correspond to solid material and white regions to void space.

%{\bf Possible future work directions:} The reader may be interested in exploring larger computational domains and confirming that the findings reported in what follows still apply; another direction of interest may be the exploration of the variability of the features discussed next across different realizations.

\subsection{Feature Extraction and descriptor construction for Machine Learning (ML)}
\label{sec:feature}

 ML is a powerful research tool, which can automatically detect patterns in complex data and make accurate predictions. It can handle large and complicated datasets; improve predictions as more data become available; and generate accurate, fast, and scalable results. To train the ML model for permeability prediction, we extract \textbf{\textit{topological, network, and structural features}} from the generated synthetic 3D porous media structures. The full list of features we consider in this work is summarized in Table~\ref{tab:feature_hierarchy}; individual features are described by category in Sections~\ref{sec:topological},  \ref{sec:networks}, and~\ref{sec:structural} in what follows. These feature descriptors are aggregated into a feature vector for model training. 

\begin{table}[thb]
\centering
\renewcommand{\arraystretch}{1.35}
\begin{tabular}{|l|l|}
\hline
\textbf{Feature Category} & \textbf{Feature Descriptors} \\
\hline

\textbf{Topological Features} &  \\ \hline

\multirow{3}{*}{Alpha Complex Approach}
& 0th dimension Total Persistence \\ \cline{2-2}
& 1st dimension Total Persistence\\ \cline{2-2}
& 2nd dimension Total Persistence\\ \hline

\multirow{3}{*}{Euclidean Distance Transform Approach}
& 0th dimension Total Persistence\\ \cline{2-2}
& 1st dimension Total Persistence\\ \cline{2-2}
& 2nd dimension Total Persistence\\ \hline

\textbf{Network Features} &  \\ \hline

\multirow{3}{*}{Network Connectivity}
& Closed Triads \\ \cline{2-2}
& Paths of Length Two\\ \cline{2-2}
& Edges\\ \hline

\multirow{4}{*}{Network Distance}
& Edge Length\\ \cline{2-2}
& Average Distance to Closest Neighbor\\ \cline{2-2}
& Average Distance to Farthest Neighbor\\ \cline{2-2}
& Closeness Centrality \\ \hline

\multirow{4}{*}{\textbf{Structural Features}}
& Sphere Diameter (PuMA) \\ \cline{2-2}
& Diffusivity \\ \cline{2-2}
& Specific Surface Area \\ \cline{2-2}
& Tortuosity \\ \hline

\end{tabular}
\caption{Summary of feature groups used in the model.}
\label{tab:feature_hierarchy}
\end{table}

\subsubsection{Topological Features}
\label{sec:topological}

Porous materials with similar structural features can conduct flow differently depending on how those features are connected, as illustrated in recent work focusing on filtration~\cite{Illingworth26}. This is where topological descriptors become essential, since they can significantly improve our ability to predict permeability in porous media, especially when using ML models. Tools from TDA, in particular PH, provide techniques for recording and quantifying topological information in datasets such as our 3D voxelized images; see~\cite{otter2017roadmap,robins2006analysis} for examples of recent work in this direction. We describe our methodology by way of examples below, first for a simple 2D dataset, before outlining how the ideas extend to complete 3D datasets.

\subsubsection*{Toy Example: 2D Dataset
\label{sec:toy}}

\begin{figure}[thb]
    \centering
    \subfigure[]
        {\includegraphics[width=0.4\linewidth]{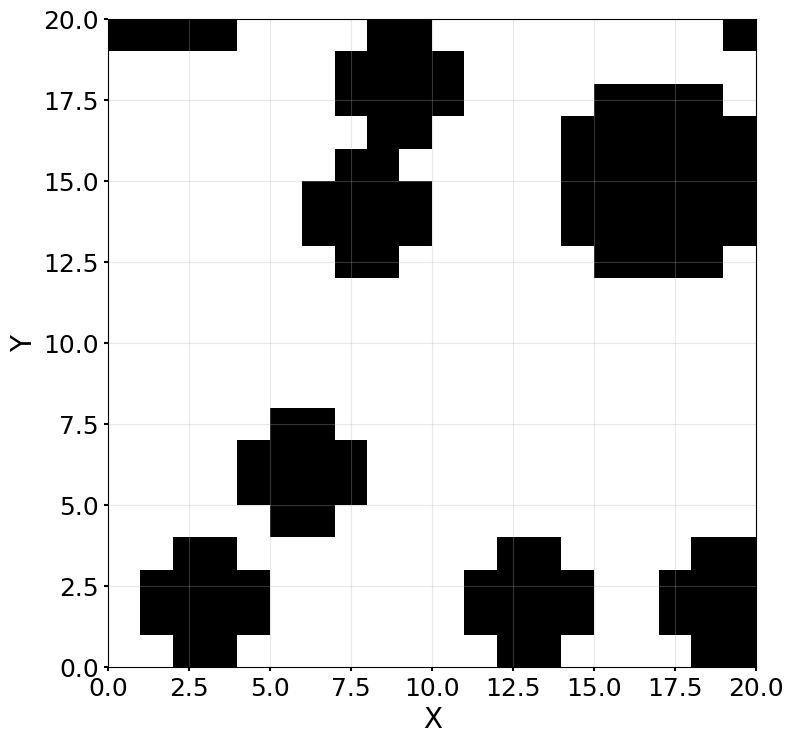}}
        \subfigure[]
        {\includegraphics[width=0.4\linewidth]{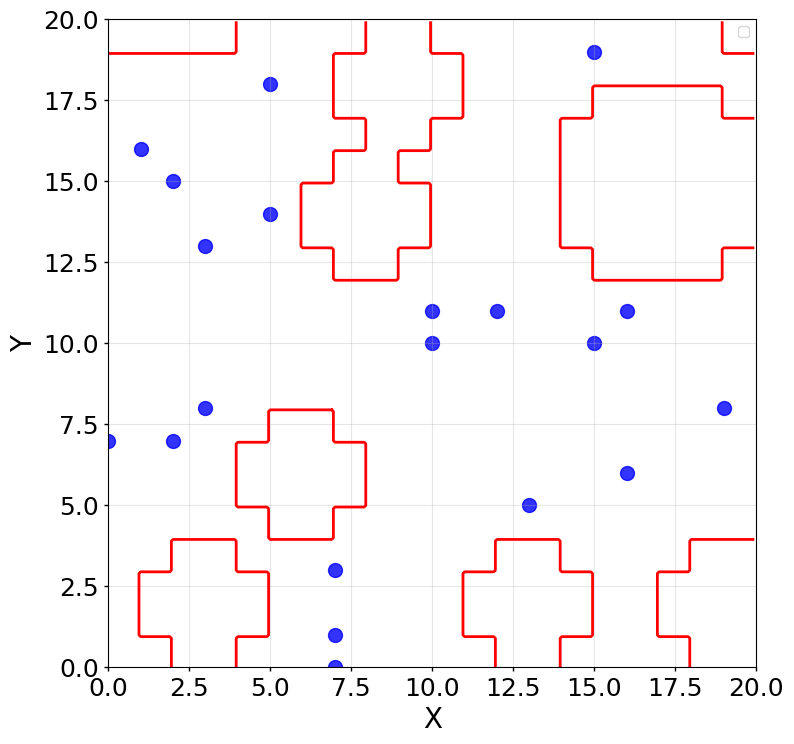}}
    \caption{(a) A void-rich 20$\times$20 2D slice perpendicular to the $z$-axis selected from a 50$\times$50$\times$50 PuMA-generated synthetic porous structure where the white regions correspond to voids and the black regions correspond to solids. (b) Void pixels are randomly subsampled to create the point cloud with 20 points.}
    \label{fig:point cloud.png}
\end{figure}

We consider a simplified domain to illustrate the basic ideas of PH that allow us to quantify the topological features in complex structures. We select a 2D $20\times 20$ slice perpendicular to the $z$-axis from a $50\times 50\times 50$ synthetic 3D porous medium generated using PuMA as shown in Fig.~\ref{fig:point cloud.png}(a). In the selected 2D image, the white region represents the voids, and the black region the solid material. We note that both in this toy example and for the larger 3D datasets considered later, we assume periodic boundary conditions, which play a role in particular when computing topological measures.

The first approach we describe involves \textbf{\textit{Alpha ($\alpha$) Complexes}}~\cite{edelsbrunner1994threedimensionalalphashapes}. Alpha Complexes are structures built from a point cloud by connecting points that are close together, controlled by a scale parameter $\alpha$. Let us explain this step-by-step. 

First, we create a point cloud from the 2D slice by subsampling the void region. Specifically, rather than using all void pixels, we randomly sample a fixed number of void pixels and place exactly one point at a random location inside each selected pixel. Random subsampling avoids bias from a regular grid and reduces data size, making TDA less computationally intensive. In this example, we sample 20 random void pixels from the slice; see Fig.~\ref{fig:point cloud.png}(b).

The $\alpha$-complexes are constructed by drawing a ball (circle in 2D) of radius $\alpha$ around each point in the point cloud. As shown in supplementary animation 1, at very small $\alpha$, there are many disconnected balls. As $\alpha$ increases, the balls become larger: when two balls overlap, edges are formed between the points. As $\alpha$ continues to increase further, and when at least three balls overlap, loops may appear. Consequently, from 2D images, we obtain zero-dimensional features, which are connected components, and one-dimensional features, which are loops; these concepts are discussed further below. The geometric structures that we build from our data points at a given scale of $\alpha$ are called $\alpha$-complexes. We construct $\alpha$-complexes and derived measures using GUDHI (Geometry Understanding in Higher Dimensions)~\cite{gudhi}, an open-source C++ library with a Python interface. 

\begin{figure}[tbp]
    \centering
    \subfigure[]
        {\includegraphics[width=0.4\textwidth]{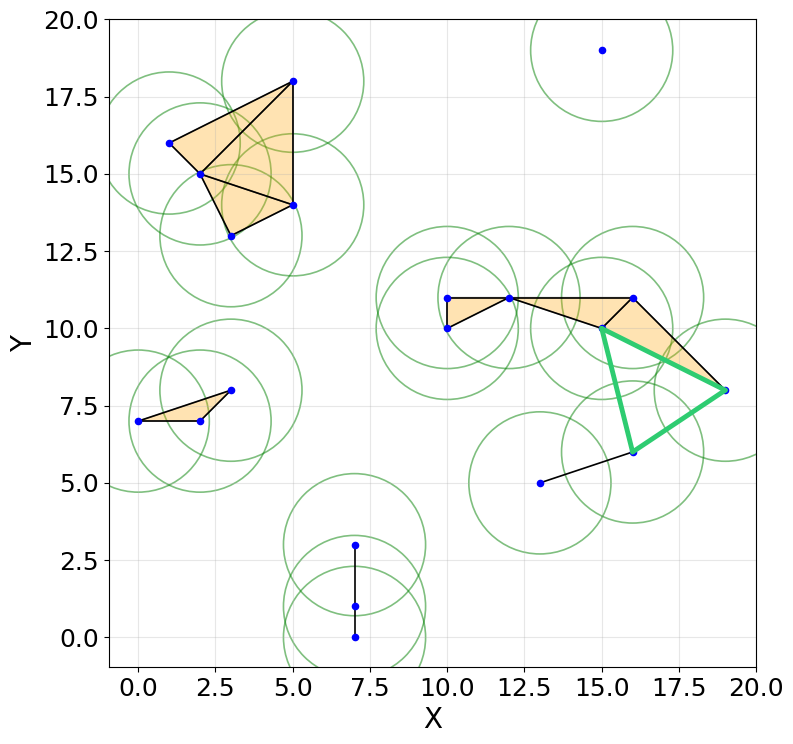}}
\subfigure[]
        {\includegraphics[width = 0.4 \textwidth]{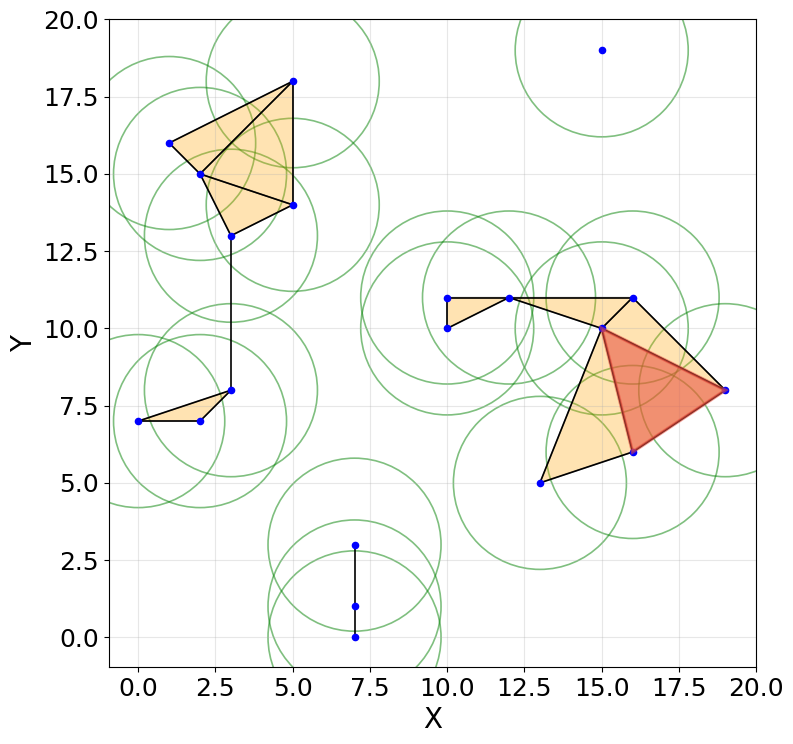}}
        
    \caption{(a) Visualization of a loop born at $\alpha$ = 2.1 highlighted in green. (b) At $\alpha$ = 2.3, the loop is filled with red, which means the region is filled and the loop disappears (dies).}
    \label{fig:alpha complex_h0.png }
    \label{fig:alpha_loop}
\end{figure}

Figure~\ref{fig:alpha complex_h0.png}(a) shows the $\alpha$-complex formation at $\alpha$ = $0.7$. All points are born at $\alpha$ = 0.0, and each point is then considered an isolated connected component. As $\alpha$ increases to 0.7, two balls overlap (in fact, two such events occur simultaneously); an edge forms between the points, and this marks the death of one of the corresponding connected components.  As $\alpha$ increases, circles around points begin to overlap, and when at least three circles overlap and the middle is empty, a loop is formed. When $\alpha$ becomes large enough that the three circles have a common intersection region, the $\alpha$-complex adds the filled triangle, and the loop dies. In Fig.~\ref{fig:alpha complex_h0.png}(b), at $\alpha$ = $1.3$,  the three edges and the filled triangle enter at the same value, and the filled loop is visible at this instant. Hence, this loop dies as soon as it is born. Figure~\ref{fig:alpha_loop} illustrates a more persistent loop, which appears at $\alpha=2.1$, where three circles overlap forming a triangle with an empty middle region highlighted in green (Fig.~\ref{fig:alpha_loop}(a)). When $\alpha$ increases to 2.3, this empty region closes and disappears; the triangle is filled (marked with red color in Fig.~\ref{fig:alpha_loop}(b)), and the loop dies.

\begin{figure}[tbp]
\subfigure[]
        {\includegraphics[width=0.4\linewidth]{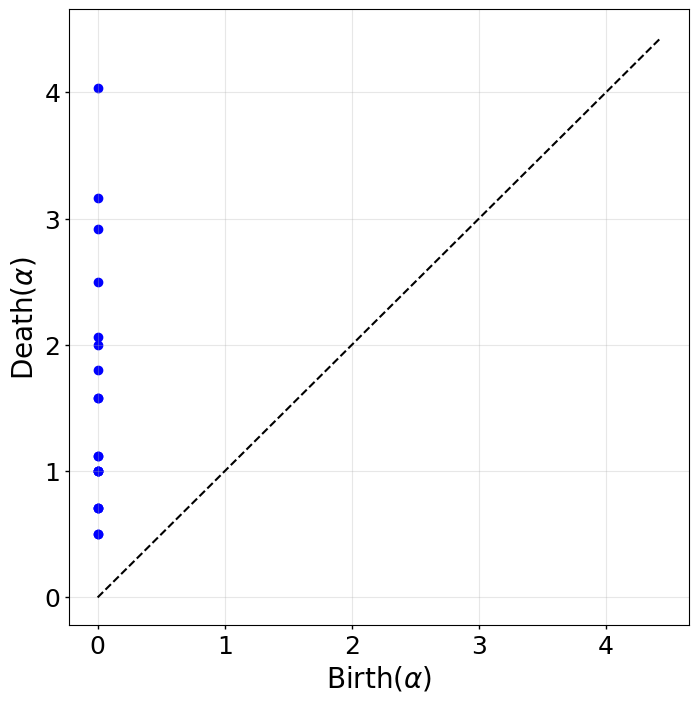}}
\subfigure[]
        {\includegraphics[width=0.4\linewidth]{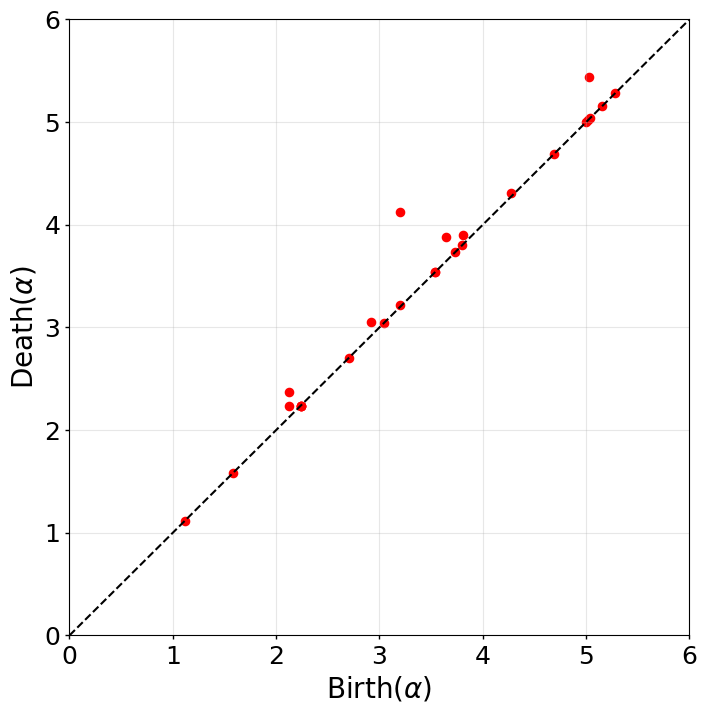}}
        \caption{Persistence diagrams (PDs) resulting from $\alpha$- complexes computed using the GUDHI library for the point cloud of Fig.~\ref{fig:point cloud.png}(b). (a) The dimension-0 diagram shows the birth and death of connected components as $\alpha$ increases. (b) The dimension-1 diagram shows the birth and death of loops as $\alpha$ increases.}
    \label{fig:pd_GUDHI.png}
\end{figure}

The appearance of a feature at a particular value of $\alpha$ marks its birth, and the disappearance of that feature as we increase $\alpha$ marks its death. The difference between death and birth values gives the lifespan of a feature. These observations allow us to plot \textbf{\textit{Persistence Diagrams (PDs)}}, which are topological summaries that plot birth vs death for each topological feature.  The dimension 0 and dimension 1 PDs for the $20\times 20$ 2D slice of Fig.~\ref{fig:point cloud.png} resulting from $\alpha$-complexes, obtained using GUDHI, are shown in Fig.~\ref{fig:pd_GUDHI.png}, which we now explain with reference to Figs.~\ref{fig:alpha complex_h0.png} and~\ref{fig:alpha_loop}. The implicit assumption is that, since points are sampled randomly from the void space, provided we take a sufficiently large sample, the PH of the point cloud should be representative of that of the void space.

\begin{figure}[tbp]
    \centering
\subfigure[]
        {\includegraphics[width=0.4\linewidth]{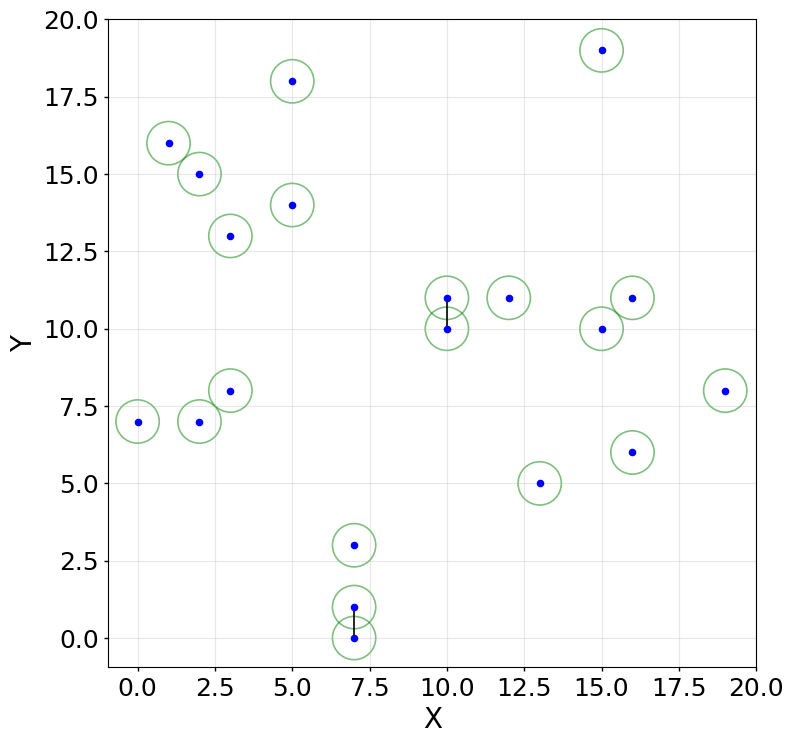}}
\subfigure[]
        {\includegraphics[width=0.4\linewidth]{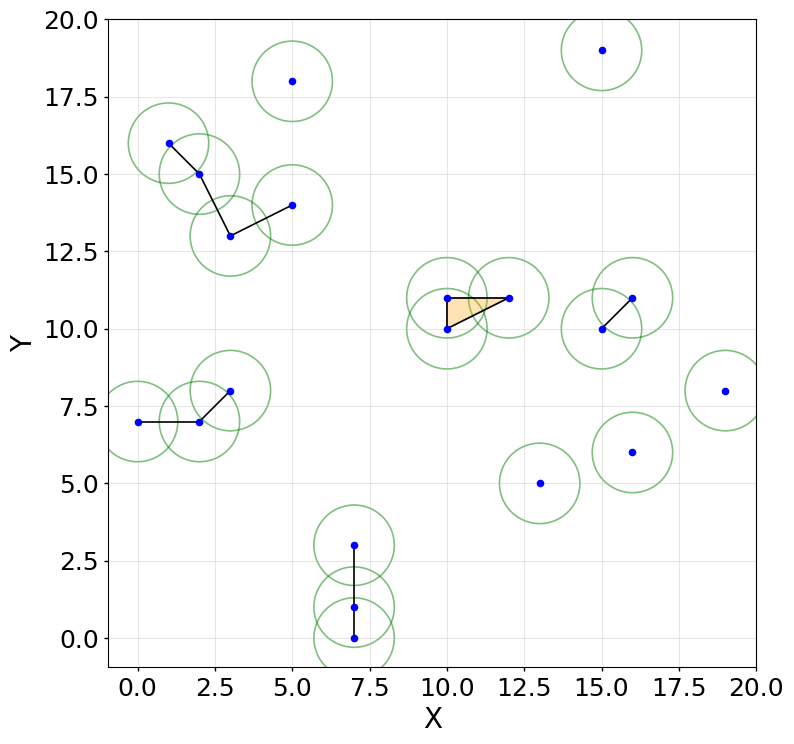}}
    \caption{$\alpha$-complex formation when circles of a particular radius $\alpha$ are drawn around each point in the point cloud. Edges are formed when two circles overlap. When at least three circles overlap at a common point, loops appear.
    (a) $\alpha$-complex formation at $\alpha$ = 0.7  (b) $\alpha$-complex formation at $\alpha$ = 1.3.}
    \label{fig:alpha complex_h0.png}
\end{figure}

The dimension 0 PD describes the birth and death of connected components as $\alpha$ varies. 
For example, we see a blue point at $(0.0,0.7)$ in the PD of dimension 0, Fig.~\ref{fig:pd_GUDHI.png}(a), marking the birth and death coordinates of these features (in fact, there are two indistinguishable points since, as noted above, there are two such events). Likewise, as $\alpha$ continues to increase, the deaths of other connected components occur, and are similarly marked by blue dots in Fig.~\ref{fig:pd_GUDHI.png}(a); see supplementary animation 1 for additional insight (all birth numbers are zero as noted above). 

The dimension 1 PD describes the birth and death of loops. Consider, for example, the red point with coordinates $(2.1, 2.3)$ in Fig.~\ref{fig:pd_GUDHI.png}(b). This represents the loop of $\alpha$-complexes born at $\alpha =2.1$ and existing until $\alpha=2.3$, described alongside Fig.~\ref{fig:alpha_loop}.  Points close to the diagonal in the PD have their birth and death coordinates approximately equal. This means that the features they represent exist only very transiently, dying almost as soon as they are born, and are thus topologically insignificant. Points farther from the diagonal have longer lifespans and represent persistent, meaningful features. Supplementary animation 1 offers an intuitive view of these features, facilitating understanding of how the topology evolves as $\alpha$ increases.

% The reader may be interested in exploring whether considering sampling of solid instead of void space provides complementary or additional information regarding the topology of a considered data set.

\begin{figure}[tbp]
    \centering
    \subfigure[]
        {\includegraphics[width=0.4\textwidth]{images/2D_slice.png}}
\subfigure[]
        {\includegraphics[width = 0.45 \textwidth]{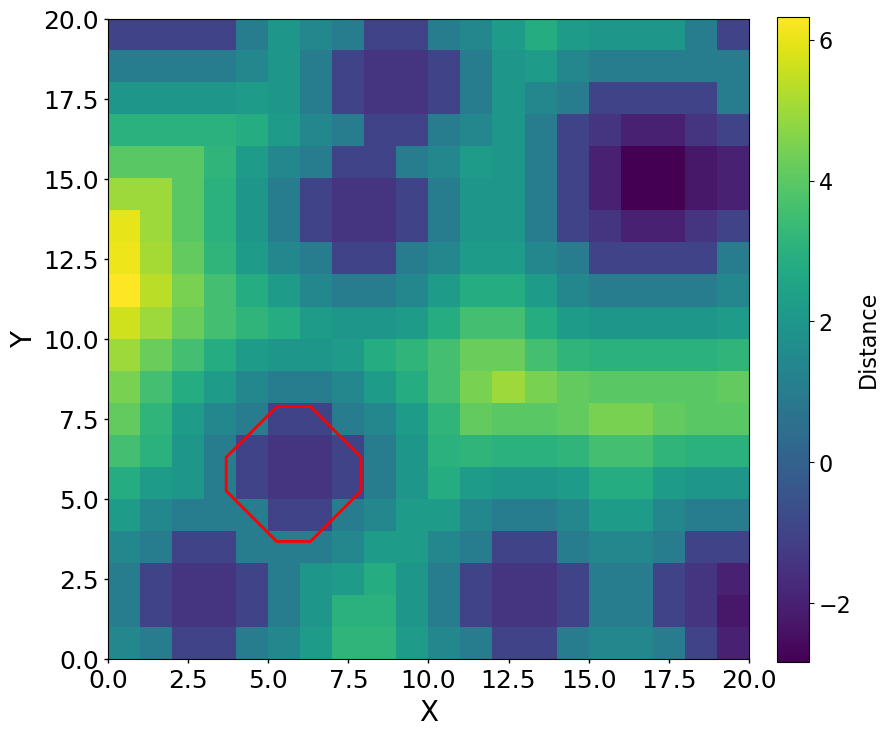}}
        
    \caption{(a) The 2D sample of Fig.~\ref{fig:point cloud.png}(a); and (b) the corresponding Euclidean Distance Transform, where each void pixel is assigned its signed Euclidean distance to the nearest solid boundary. Colors indicate distance magnitude, with yellow/green showing pixels deep inside pores and purple/blue showing pixels inside the solid region. The single loop formed at threshold level 1 is highlighted in red in Fig.~\ref{fig:EDT_new}b. }
    \label{fig:EDT_new}
\end{figure}

\begin{figure}[tbp]
    \subfigure[]
        {\includegraphics[width=0.39\linewidth]{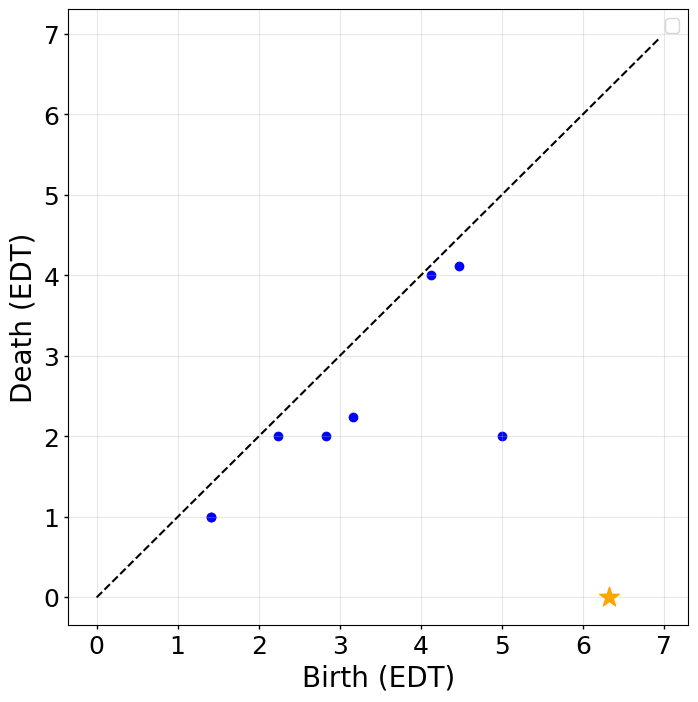}}
        \subfigure[]
        {\includegraphics[width=0.4\linewidth]{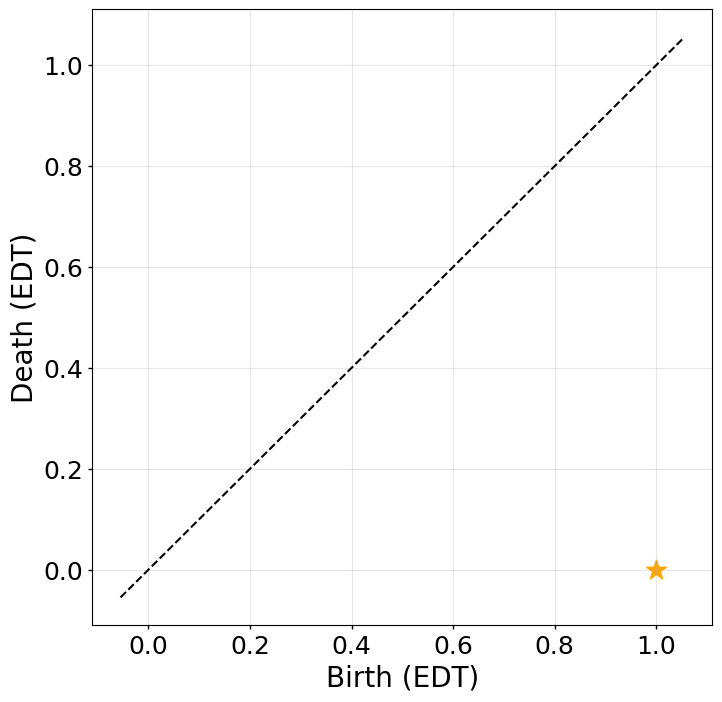}}
    \caption{ Persistence diagrams (PDs) resulting from EDT (a) the dimension-0, and  (b) the dimension-1. The components or loops that appear first (at the highest threshold value) are represented by yellow stars. }
    \label{fig:pd_EDT}
\end{figure}

The \textbf{\textit{Euclidean Distance Transform (EDT)}} \cite{suzuki2021flow} is another method that we use to extract topological features using PH. We use the same 2D $20\times 20$ slice shown in  Fig.~\ref{fig:point cloud.png}(a) to illustrate the concept.  Each pixel in the void space is assigned a value equal to its signed Euclidean distance from the nearest solid boundary. Pixels deep inside the void space have large positive values, while those near walls have smaller values, and those inside solid regions are associated with negative values, as shown in Fig.~\ref{fig:EDT_new}.  The zero contour delineates the solid–void boundary.   Figure~\ref{fig:EDT_new}(a) shows the original 20$\times$20 2D slice and Fig.~\ref{fig:EDT_new}(b) shows its corresponding color-coded EDT, with a color scale representing the Euclidean distance from each image pixel to the nearest solid boundary. With the chosen color scale, yellow/green indicates pixels farther from the solid inside the void region (large positive values), and purple/blue indicates pixels inside the solid region (negative values). These EDT values are used as input to HomCloud~\cite{HomCloud}, a package that we use for the computations. 

We perform superlevel filtration on the EDT data, starting from the maximum distance and gradually decreasing the threshold to zero (since we are analyzing the topology of the void region, we do not consider negative values). The superlevel filtration can be visualized like a fog descending on a landscape; first, the highest hilltops appear above the fog, forming isolated points (dimension-0 features). As the threshold decreases, more pixels are included, and previously disconnected components may connect, forming loops (dimension-1 features). 

Initially, only pixels with large distance values (points located deep within the void) are included. As we decrease the threshold, more pixels closer to the solid boundaries appear. When two dimension-0 connected components merge, the later-born component dies, while the older component persists. The process is most clearly seen in Supplementary animation 2. For the connected components (dimension-0 features), one can easily discern one isolated bright area in Fig.~\ref{fig:EDT_new}(b) near $x=0,\, y=11.5$, that is born at the highest EDT value of about 6.32 and never dies. For graphical purposes,  we set its death EDT value to zero. This feature is shown in Fig.~\ref{fig:pd_EDT}(a) by a yellow star. Decreasing the threshold further, we reach the area around  $12.5<x<20, y=8$ that produced the generators in Fig.~\ref{fig:pd_EDT}(a) with the birth coordinate larger than four. Two of the components persist only briefly before merging with the older one and are therefore very close to the diagonal. Other birth-death events in Fig.~\ref{fig:pd_EDT}(a) may be similarly explained with the aid of Supplementary Animation 2.

Homcloud does not assume periodic boundary conditions; hence, the domain boundaries are treated as physical boundaries. Loops cannot cross these domain boundaries. Therefore, only one loop is formed in this case, at EDT threshold level 1, which is marked in red in Fig.~\ref{fig:EDT_new}(b). A careful examination of Supplementary Animation 2 confirms the appearance of this single loop when the EDT threshold is decreased to 1; consistently with the dimension-0 discussion, that loop never dies, and for the simple dataset considered here, no more loops form.  

%{\bf Possible future work directions:} There are numerous options for an alternative approach to the computation of EDT. The reader may be interested in exploring (i) whether considering negative EDT values provides additional information, or (ii) confirming that the expected connection between sub- and superlevel filtration holds.

\subsubsection*{Extension to 3D data\label{sec:3D-TFE}}

The above example introduces the key PH concepts describing topology for a simple toy 2D image. These concepts extend naturally to the 3D porous structures that we generate with PuMA by implementing the two methods just discussed.  In 3D, PH captures an additional class of 3D topological features corresponding to enclosed cavities (dimension-2), in addition to connected components (dimension-0) and loops (dimension-1). Aside from the additional PD for dimension-2 features, the process is the same as outlined in the 2D toy example.  For simplicity, when calculating $\alpha$-complexes, we do not use subsampling, but instead place one point inside each void voxel.   
 
The next step is to use the topological information derived from these PDs to train the ML model, which requires converting the PDs into scalar measures suitable for ML. One such scalar measure is the Total Persistence (TP), calculated by summing the lifetimes (death minus birth) of all topological features in a given PD. Thus, each PD is aggregated into a scalar value (TP), yielding topological descriptors for the ML model.  We note that, in calculating TP, Homcloud~\cite{HomCloud} includes only topological features with finite birth and death counts. The features that appear first and never disappear are not included in TP computations.

Our preliminary results (not reported for brevity) suggest that, as expected, subsampling in the context of $\alpha$-complexes reduces computing time significantly.  One may want to consider developing appropriate measures that can quantify the appropriate degree of subsampling, or in other words, answer the following question: What is the minimum level of sampling that quantifies with acceptable accuracy the topology of the considered data set?  Furthermore, for both $\alpha$-complexes and the EDT, it may be of interest to explore other ways of vectorizing PH  beyond the simple Total Persistence measure; see~\cite{bubenik_jsc_2017} for an approach based on persistence landscapes, and~\cite{ali2023} for a recent review. We leave these directions for future work.

\subsubsection{Network features}
\label{sec:networks}
Reducing the interconnected pore structure to a representative network provides a compact, physically interpretable representation of the flow pathways that determine the permeability of a porous structure. Such networks are intended to capture the essential connectivity of a porous structure in a simplified graph, thereby reducing the complexity of the underlying voxel-based porous media dataset. We extract the networks from a digitized representation of a porous medium using the SNOW2 algorithm implemented in PoreSpy \cite{porespy}. 

 We again illustrate the concept using the $20\times20$ 2D slice (shown in Fig.~\ref{fig:network_2d}(a)) that we used previously. The SNOW2 algorithm applied to this 2D slice yields the network representation shown in Fig.~\ref{fig:network_2d}(b), which shows the nodes (void space centers) and the throats connecting adjacent nodes, with different colors representing the void regions as discussed in what follows. 
 
 While the full SNOW2 implementation is detailed in the PoreSpy documentation, we briefly outline the core steps below for clarity, with reference to the 2D slice shown in Fig.~\ref{fig:network_2d}(a):
\begin{figure}[tbp]
    \centering
\subfigure[]
        {\includegraphics[width=0.4\linewidth]{images/2D_slice.png}}
\subfigure[]
        {\includegraphics[width=0.4\linewidth]{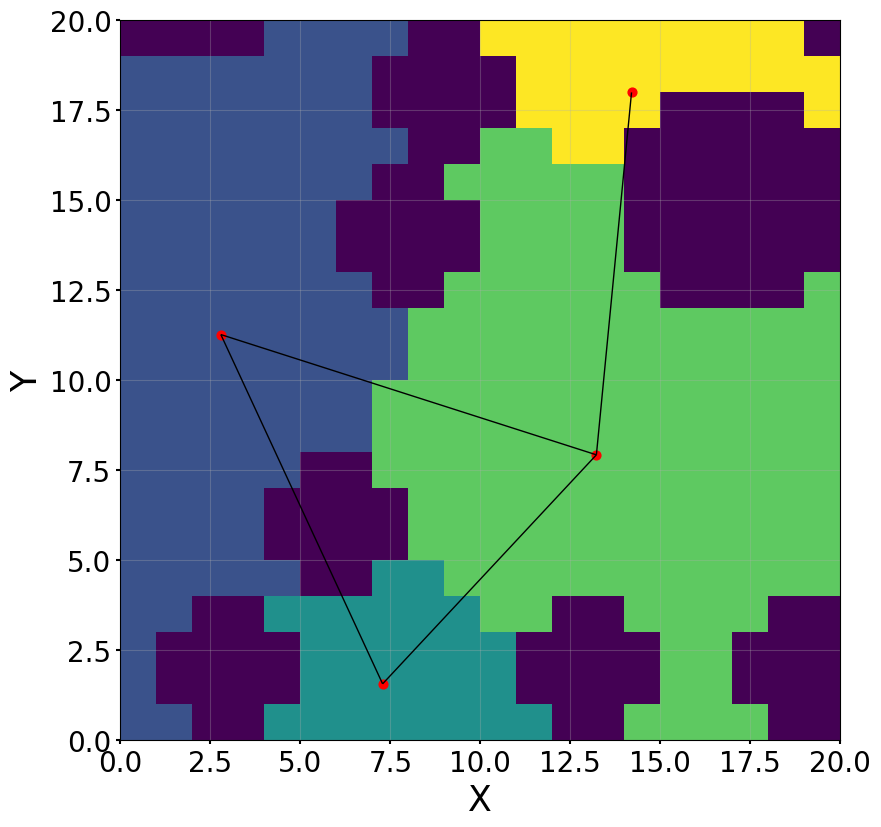}}
        \caption{ (a) The 2D sample of Fig.~\ref{fig:point cloud.png}(a); and (b) the corresponding network obtained using the SNOW2 algorithm in PoreSpy~\cite{porespy}, where nodes correspond to void space centers and edges represent throats connecting adjacent nodes. The different colors represent the disjoint void regions described in Step 4 of the algorithm.}
    \label{fig:network_2d}
\end{figure}

\begin{figure}[tbp]
    \centering
    \subfigure[]
        {\includegraphics[width=0.4\linewidth]{images/3D_structure.png}}
        \subfigure[]
        {\includegraphics[width=0.4\linewidth]{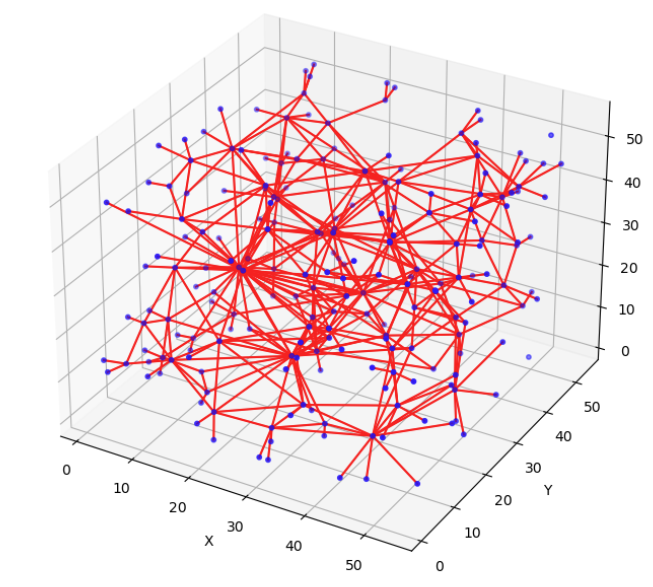}}
\caption{\label{fig:Porespy.png} (a) 3D visualization of a typical synthetic porous medium dataset of $50 \times 50 \times 50$ voxels generated using PuMA~\cite{puma} software. (b) Network Extraction of the PuMA-generated synthetic porous structure using PoreSpy~\cite{porespy}, where nodes correspond to 3D hole centers and edges represent throats connecting adjacent nodes.}
\end{figure}

\paragraph{\textbf{Step 1}}
The algorithm creates a distance map, computing the Euclidean Distance Transform (EDT), which assigns to each point in the void space the Euclidean distance to the nearest solid wall. Points deep inside a void space have higher EDT values, while points near a wall have lower ones. 
However, artifacts can appear on this map, creating artificial local maxima in EDT values that do not correspond to true void-space centers. To fix this, the method applies a blur filter to smooth the surface.

\paragraph{\textbf{Step 2}}
The second step is to find the peaks in the distance map, which represent the centers of void space (the points farthest from the walls). These peaks become the nodes in the network representation. The algorithm accomplishes this by examining a specific neighborhood around each pixel. If the search area is too small, the algorithm might find false peaks; if it is too large, it could miss some real ones.  SNOW2 chooses an appropriate search radius that correctly identifies the true void space centers based on the distance map.

\paragraph{\textbf{Step 3}}
The next step is to remove redundant peaks from the set found in Step 2. The algorithm checks all detected peaks to remove duplicates and false positives by slightly expanding each peak and comparing them to see if they overlap. If two peaks are close together and represent the same void space centers, they are merged into one. By default, the algorithm keeps the peak farthest from the wall.

\paragraph{\textbf{Step 4}}
The final step of the algorithm is to segment the image into disjoint void-space regions using the watershed algorithm, which, for the present purposes, could be imagined as “flooding” the structure from each void-space center. As the flooding liquid fills the space, it spreads until it meets the liquid from another void-space center. An edge (throat) is created between two nodes only if their corresponding regions touch in the segmented image. This produces a 2D network in which nodes represent centers of void space and edges represent throat connections between neighboring centers. 

Figure~\ref{fig:network_2d}(b) shows the outcome of the SNOW2 algorithm on the $20\times 20$ 2D slice; we note that, while the main features of the sample are well captured, the algorithm is far from perfect; in particular, we note a throat that passes through the solid region, showing that the considered simplified description of a porous medium by a network comes at an accuracy cost.  

Extending to 3D, Fig.~\ref{fig:Porespy.png}(a) shows the 3D visualization of a typical synthetic porous medium dataset of $50 \times 50 \times 50$ voxels generated using PuMA software, and Fig.~\ref{fig:Porespy.png}(b) shows the corresponding network extracted using PoreSpy.

To summarize, the SNOW2 algorithm provides a simplified, approximate network representation of the medium under consideration. Once such a description is available, we may quantify it using network-based descriptors, which are then fed to the neural network. We separate these descriptors into two groups: network connectivity descriptors (the first three items below) and network distance descriptors (the subsequent four items). A single scalar value is recorded for each descriptor for each network.
\begin{itemize}
    \item \textbf{\textit{Closed Triads:}} the total number of closed triads, defined as sets of three nodes that are all mutually connected.
    \item \textbf{\textit{Paths of Length Two:}} any set of three nodes where the first node is connected to a middle node, and the middle node is connected to a third node. Note that these triplets may also be part of longer paths in the network.
    \item \textbf{\textit{Edges:}} the total number of direct connections between pairs of nodes in the network.
    \item \textbf{\textit{Edge Length:}} the mean Euclidean distance between two connected nodes.
    \item \textbf{\textit{Average Distance to Closest Neighbor:}} the average distance from a node to its nearest connected neighbor.
    \item \textbf{\textit{Average Distance to Farthest Neighbor:}} the average distance from a node to its most distant connected neighbor.
    \item \textbf{\textit{Closeness Centrality:}} quantifies how close a node is to all other nodes in a system. It is defined as the average of the reciprocal sum of the shortest paths between individual nodes. High centrality values correlate with shorter paths to other nodes.
\end{itemize}

\subsubsection{Structural Features}
\label{sec:structural}

We refer to measurable characteristics of the material’s internal void space geometry as structural features. They describe how the void spaces and solid spheres are arranged, shaped, and connected in our raw voxelized data.  The structural features we extract are the sphere diameters, the diffusivity of the void space, the specific surface area, and the tortuosity of void paths:\\
    $\bullet$ \textbf{\textit{Diameter (of random sphere structure):}} diameter of the spheres used during the random sphere generation process in PuMA, which is set here to 5 - 15 voxels. For each structure, a single diameter value between 5 and 15 voxels is randomly selected and used for all spheres within that structure. Thus, diameter is represented by a single scalar value for each structure.\\
    $\bullet$ \textbf{\textit{Diffusivity:}} a measure of how easily molecules or particles can move or diffuse through the void space. Effective diffusivity of the void space is computed using PuMA's continuum tortuosity solver. Diffusivities are calculated separately in the x-, y-, and z-directions and summed to obtain a single scalar feature for each structure.\\
    $\bullet$ \textbf{\textit{Specific Surface Area:}} indicates the amount of internal surface area of the void space. PuMA calculates the internal surface area per unit volume of the structure and is recorded as a single scalar value for each sample.\\
    $\bullet$ \textbf{\textit{Tortuosity:}} measures how winding or indirect the pathways through the void space are. Effective tortuosity is computed using PuMA's continuum tortuosity solver. Tortuosities are calculated independently in the x-, y-, and z-directions and summed to obtain a single scalar feature for each structure.
    
  All these are computed in PuMA~\cite{puma} using the respective built-in functions. These structural descriptors, which are summarized in Table~\ref{tab:feature_hierarchy}, are then fed into the neural network.

\subsection{Ground-Truth Permeability Simulation}
\label{sec:permeability}

Permeability for each synthetic 3D porous sample is calculated using PuMA~\cite{puma}, which solves the Stokes equations to simulate fluid flow through the void space. Each 3D structure is used as a simulation domain, and a pressure difference is applied across it. From the resulting fluid motion, the effective permeability tensor of the structure is obtained. The diagonal components of the permeability tensor are averaged and converted to Darcy units (a unit of permeability where $1$ Darcy is equivalent to $9.87 \times 10^{-13}\,\mathrm{m}^2$) to obtain a single permeability value for each sample. These are the ground-truth permeability values used to train the machine-learning (ML) model discussed next.

\subsection{Machine Learning: Neural Network Model}
\label{sec:ML}

An ML model learns patterns from data and predicts permeability rather than being given an explicit equation; for this purpose, we allow the model to discover this relationship by analyzing many examples. We implement neural network-based ML to predict the permeability of porous materials, offering a fast, efficient alternative to traditional simulation methods. A neural network consists of layers of interconnected “neurons” that detect and learn from complex patterns in the data. Neural networks are powerful because they can represent and learn complex relationships between inputs and outputs that traditional models, such as linear regression and random forests, cannot capture; we note that a similar ML approach to the one outlined here was used recently by R\"oding et al.~\cite{roding2020}. The network has one input layer, three dense layers of 128 nodes each, and one output layer. In total, this network structure creates approximately 50,000 parameters for the model to tune, providing ample freedom for it to learn to predict permeability. The algorithm is implemented in TensorFlow~\cite{tensorflow2015-whitepaper}. 

We represent each porous structure by a single fixed-length feature vector by concatenating structural, network, and topological information, and labeling it with its corresponding permeability calculated from PuMA, creating a complete dataset. We split our 1,000-structure dataset into 700 training samples, 150 validation samples, and 150 test samples. Here, the training dataset is used to train our model to predict permeability. Each node in the layer takes the input numbers (in our case, topological,  network, and structural features of the porous structure), multiplies them by adjustable numbers called weights, adds a small correction called bias, and produces an output. The weights and biases are the model parameters, which determine how strongly each feature influences the predicted permeability. During training, the network makes a prediction and compares it to the true permeability value (computed using PuMA). The difference between the predicted and true values is quantified using a loss function, for which we use the mean squared error (MSE), measuring the average squared difference between prediction and truth. If the error is large, the model adjusts its parameters to reduce it. This adjustment is performed using the Adam (Adaptive Moment Estimation)  optimizer~\cite{reyad2023modifiedadam}, which efficiently updates the weights and biases to improve accuracy.

After each node performs its calculation, the result passes through an activation function. We use the Rectified Linear Unit (ReLU) activation function~\cite{math14010039}, defined as $f(x)=max(0,x)$,
 which outputs the input directly if it is positive, and zero otherwise. ReLU introduces nonlinearity, allowing the network to learn complex relationships between structure and permeability. Without this step, the model could only learn simple linear relationships.

To train the model efficiently, the dataset is divided into small groups called batches. We use a batch size of 32, meaning that network parameters for 32 porous structures are updated at a time. When the model has processed all training samples once, we call it one epoch. We train the model for 1,000 epochs, the number chosen empirically, as discussed below. As training progresses, prediction accuracy for the training data improves. If the model is trained for too long, however, it may start memorizing the training data instead of learning general patterns, a phenomenon known as overfitting. When overfitting occurs, the model may perform very well on the training dataset, but not as well on new, unseen data. To monitor this, we use a separate validation set (150 structures) to ensure the model maintains strong predictive performance beyond the training set. The validation set provides an unbiased measure of the model’s predictive ability: since the model has not seen these samples during training, good performance on this set indicates that the model has learned general patterns rather than simply memorizing the training data. Once trained, our neural network can rapidly predict permeability for new porous structures, providing an efficient alternative to computationally expensive full numerical simulations. 

\section{Results and Discussions}
\label{sec:results}

In this Section, we evaluate the accuracy and effectiveness of the proposed model for various sets of input features. In the following, let $k$ denote the true permeability obtained from the PuMA simulations; $\hat{k}$ represents the predicted permeability. 

\begin{figure}[tbp]
    \centering
\subfigure[]
        {\includegraphics[width=0.4\linewidth]{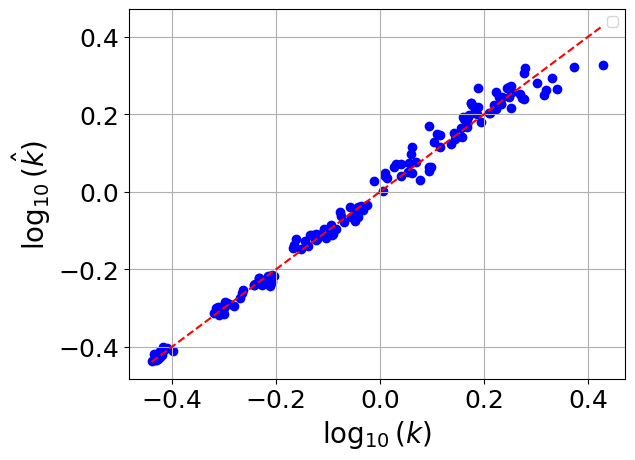}}
    \subfigure[]
        {\includegraphics[width=0.4\linewidth]{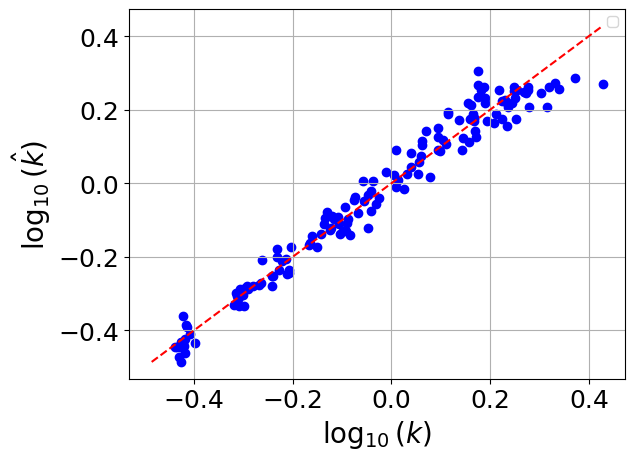}}
    \caption{Predicted permeability values $\log_{10}(\hat{k})$ vs the true permeability values $\log_{10}(k)$
for the neural network model: (a) using all the descriptors;  (b)  using only the network connectivity descriptors. }
    \label{fig:predicted_vs_new.png}
\end{figure}

\textbf{\textit{Predicted vs True Permeability:}}
Figure~\ref{fig:predicted_vs_new.png} compares the logarithmic values of the predicted permeability, $\log_{10}(\hat{k})$, with the true permeability, $\log_{10}(k)$, for the neural network model. We use the logarithmic values of permeability because they span several orders of magnitude, simplifying the comparison. The red dashed line on which $\log_{10}(\hat{k})=\log_{10}(k)$ represents perfect agreement between the predicted and true values.
  
When all structural, network, and topological descriptors are used as input features, most data points lie close to the red dashed reference line (as shown in Fig.~\ref{fig:predicted_vs_new.png}(a)), indicating excellent agreement between predictions and ground-truth values. This observation is supported by a coefficient of determination of $r^2 = 0.9834$.  In contrast, when only network-connectivity descriptors are used (as shown in Fig.~\ref{fig:predicted_vs_new.png}(b)), the data points exhibit a larger spread away from the reference line, corresponding to a lower prediction accuracy with $r^2 = 0.9665$. This example demonstrates that incorporating the full set of descriptors improves the reliability of permeability predictions compared with the subset of network-connectivity descriptors.

\textbf{\textit{Accuracy and Computation Time}}: To further evaluate the model's performance, we report several accuracy and efficiency measures: Mean Absolute Percentage Error (MAPE), Feature Extraction Time (FET), ML training time, and ML prediction time. MAPE is defined as
\begin{equation}
\mathrm{MAPE} =
\frac{100}{N_{\mathrm{test}}}
\sum_{i=1}^{N_{\mathrm{test}}}
\left|
\frac{k_i - \hat{k}_i}{k_i}
\right|,
\label{eq:19}
\end{equation}
Lower MAPE values indicate better predictive accuracy. After training, the model predicts permeability on the test set ($N_{\mathrm{test}}=150$) and MAPE is computed from those predictions only.

Table~\ref{tab:permeability_and_time} shows MAPE, FET, ML training time, and ML prediction time for eight input feature sets, with PuMA permeability included as the ground-truth reference.  FET is the time to compute the input features for 1000 structures on a machine equipped with an 11th Gen Intel(R) Core(TM) i5-1145G7 CPU @ 2.60 GHz and 16 GB of RAM. Training time is the measured neural network training time, scaled to 1000 structures. Prediction time is the time for the trained model to predict permeability, again scaled to 1000 structures. The time taken for ML model training and the time taken for permeability prediction using ML provide additional information regarding the efficiency of the ML model.

\begin{figure}[tbp]
\centering
\includegraphics[width=0.9\linewidth]{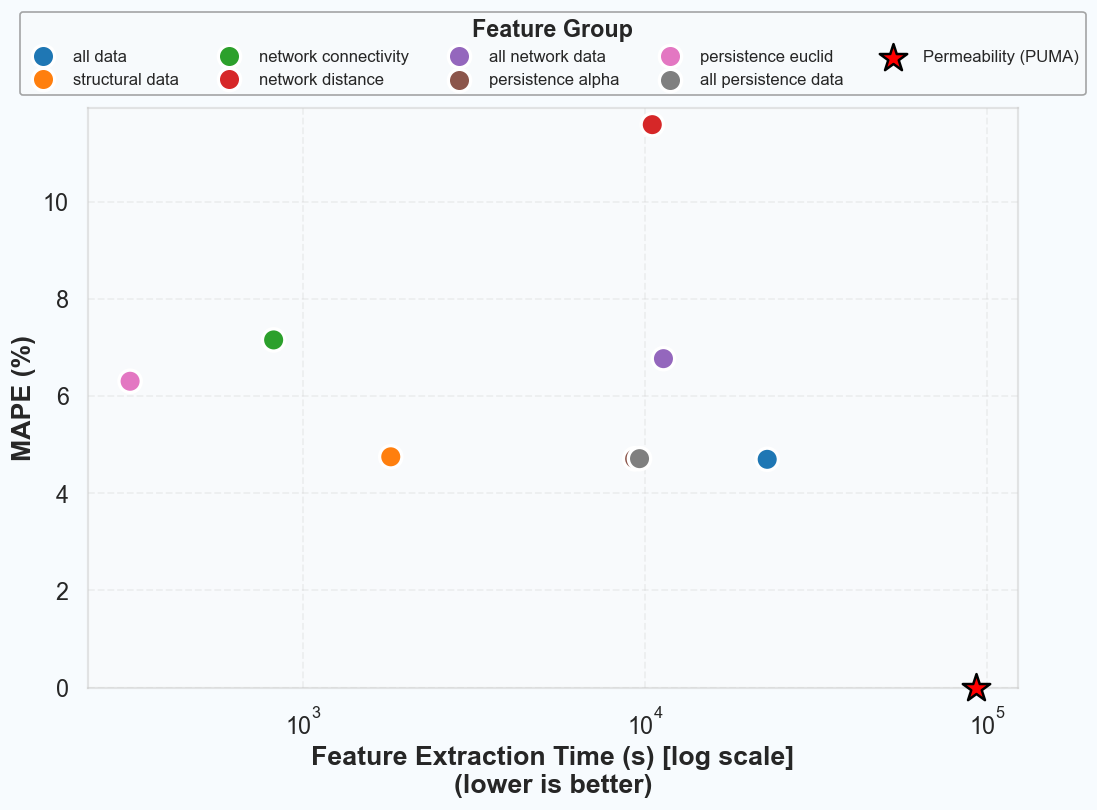}
\caption{\label{fig:Predictive Power.png} MAPE (\%) vs. Feature Extraction Time (FET) (s).  Note that `persistence alpha' and `all persistence data' are almost overlapping.
}
\end{figure}

Figure~\ref{fig:Predictive Power.png} shows the MAPE and FET results presented in Table~\ref{tab:permeability_and_time} in graphical form, illustrating the utility of the ML approach for computing permeability, as well as the trade-off between efficiency and accuracy of the various approaches considered. 

\begin{table}[thb]
\centering
\renewcommand{\arraystretch}{1.5}
\begin{tabular}{|l|c|c|c|c|}
\hline
\textbf{Feature Group} & \textbf{MAPE (\%)} & \textbf{FET (sec)} & \textbf{Training time (sec)} & \textbf{Prediction time (sec)} \\
\hline
All features & 4.70 & 22745 & 59 & 4.7 \\
\hline
Structural features & 4.75 & 1804 & 354 & 3.8 \\
\hline
Network connectivity & 7.16 & 820 & 208 & 1.6 \\
\hline
Network distance & 11.60 & 10492 & 264 & 2.1 \\
\hline
All network features & 6.78 & 11312 & 391 & 3.1 \\
\hline
Total persistence ($\alpha$-complex) & 4.71 & 9317 & 363 & 3.4 \\
\hline
Total persistence (EDT) & 6.31 & 312 & 238 & 2.8 \\
\hline
All persistence features & 4.72 & 9629 & 377 & 4.1 \\
\hline
\textbf{PuMA permeability (Ground Truth)} & \textbf{0} & \textbf{92480} & --- & --- \\
\hline
\end{tabular}
\caption{Mean Absolute Percentage Error (MAPE), Feature Extraction Time (FET), ML training time, and ML prediction time for eight input feature sets, with PuMA permeability included as the ground-truth reference.}
\label{tab:permeability_and_time}
\end{table}

Let us summarize the key takeaways of the present work:
\begin{itemize}
    \item Using all available descriptors, the model achieves a MAPE of approximately 4.70\%.
    \item  Persistence descriptors based on all persistence features achieve a MAPE of 4.72\%, while structural descriptors achieve a MAPE of 4.75\%, indicating similar, near-optimal, performance.
    \item Predictions based only on network descriptors are slightly less accurate, with a MAPE of 6.78\%. The slightly lower accuracy is expected, as the SNOW2 algorithm reduces structural complexity. However, the fact that network descriptors remain reasonably accurate indicates that SNOW2 preserves important structural information relevant for permeability prediction.
    \item Network distance-based descriptors are substantially less predictive. These features result in a high MAPE of 11.60\%, suggesting that such network-distance features contribute little useful information for permeability prediction. 
    \item The structural features are the best in terms of minimal error and time taken for feature extraction. However, they take more prediction time and training time. Network data is cheap to compute due to information reduction, but because of the simplification, the predictions are not very accurate and offer no clear advantage in training or prediction time compared to structural features. 
    \item The persistence data provide a middle ground, being relatively quick to compute and providing an accurate predictor of permeability. Note in particular that feature extraction time is significantly lower for EDT compared to $\alpha$-complexes.
    \item The PuMA calculation of ground-truth permeability of 1000 structures requires 92,480 seconds, whereas the neural network, once trained with all descriptors, completes permeability predictions of 1000 structures in 4.7 seconds with an accuracy of 4.70 \% on the test set. Thus, the ML model provides a highly efficient alternative to traditional PuMA calculations, delivering reliable permeability estimates with drastically reduced computational cost.
\end{itemize}

Before closing, we briefly discuss our results in the context of the Kozeny-Carman (KC) equation~\cite{kozeny,carman}, which is often used to obtain a quick permeability estimate. This law predicts the permeability as $k = \phi^3/(c s^2)$, where $\phi$ is the porosity, $s$ is the specific surface area (pore surface area divided by the sample volume), and $c$ is the KC constant. Since we hold porosity $\phi$ constant, this relation suggests a power-law dependence, $k \sim s^{-2}$ for the data in our study. Seeking a similar power law for our 1,000 realizations, we find (by best fit, details omitted for brevity) that $k\sim s^{-p}$ with $p \approx 1.85$, so the KC equation slightly overpredicts the dependence of $k$ on $s$. On the other hand, we may approach the problem assuming that KC holds, and find the proportionality constant $c$ by fitting our data to the KC scaling law ($p=2$, with $\phi=0.5$). When we do this, we find the MAPE of the resulting KC law to be $\rm{MAPE}_{KC} \approx 15\%$. Based on these observations, we conclude that the computed measures combined with ML provide significantly more accurate prediction of permeability than the KC prediction, even for our relatively simple setup characterized by constant porosity.   

The results presented are encouraging, providing validation of our hypothesis that data obtained using persistence diagrams for the porous structures, or from the extracted networks, are useful for predicting permeability. We find that the features we extract are, in fact, valid predictors of permeability, even if they are somewhat less powerful than the more traditional structural information that we already knew would perform well. Still, since these alternative methods are quicker to compute, we conclude that if saving computational time is important when computing permeability, using network representations and topology to predict permeability shows clear promise and offers a potential alternative to traditional simulation-based methods. 

\section{Summary and Conclusions}
\label{sec:conclusions}

In this work, we have shown that there is a significant potential for various simplified approaches to predict permeability of porous materials based on their structural properties.  In particular, we find that topological descriptors provide insightful results.  Some useful features of the topological descriptors are that they work well across different numbers of physical dimensions, are completely objective, and can be easily set up in a manner that allows for a simple interface with machine learning algorithms.   

While our results are promising, much more remains to be done in the direction of predicting performance (here, permeability) from the material structure.  Foremost, the topological approach in particular can be extended significantly in the direction of more elaborate measures.  The network-based measures also offer significant potential, as illustrated by their uses in many fields, see, e.g.~\cite{Barthelemy2011SpatialNetworks,dorogovtsev2008,Newman_2010}, which discuss in detail the type of measures used in the present work, as well as much more elaborate ones.

Another direction for future work involves incorporating a wider variety of porous structures, including experimental datasets, and exploring the use of graph neural networks to improve predictive performance. In this context we note that, while the domain size of 50$\times$50$\times$50 used here enables rapid generation of datasets and simulation within PuMA, it may not fully represent the statistical homogeneity or representative elementary volume typically required for larger porous domains. Therefore, as well as different types of sample, it will also be valuable to explore larger domains to assess how sample size affects the results.  We see this particular aspect, including also a careful analysis of how computation time scales with the domain size, as a natural extension of the present work. 

\acknowledgments
We thank the members of the Spring 2025 Applied Mathematics Capstone class for their contributions, discussions, and collaborative efforts throughout the semester: Madison Abella, Kouadio Assale, Alexander DeFilippo, Ethan Diaz, Nick Hand, Amiri Hayes, Faiza Iqbal, Bryan Kline, Carlos Layme, Joey Mucci, and Cailyn Stile. This work was supported by the grants NSF DMR 2410985, DMS 2201627, ACS PRF 69100-ND9, and the NJIT GHAIRI Seed Program. 
\bibliographystyle{unsrt}
\bibliography{soft_matter}
\end{document}

\title{Topological Data Analysis combined with Machine Learning for Predicting Permeability of Porous Media}

\author{E. Dagdelen$^{1}$$^{*}$, C.N. Lalu $^{2}$$^{*}$, A. Karlekar$^{1}$,  M. Arora $^{1}$, M. Illingworth$^{1}$, J. Jaquette $^{1}$, L.J. Cummings$^{1}$ and L. Kondic$^{1}$}
%\thanks{}
\affiliation{$^{1}$ Department of Mathematical Sciences, New Jersey Institute of Technology, Newark, New Jersey 07102, USA\\
$^{2}$ Department of Physics, New Jersey Institute of Technology, Newark, New Jersey 07102, USA
$^{*}$ First and second authors contributed equally.}

\maketitle

\section*{List of supplementary files:\\}
\vskip 1in

\paragraph{Video 1:} Evolution of the $\alpha$-complex as $\alpha$ increases, showing the formation and disappearance of connected components (H0) and loops (H1). Circles of a particular radius $\alpha$ are drawn around each point in the point cloud. Edges are formed when two circles overlap, and when at least three circles overlap at a common point, loops appear.\\

\paragraph{Video 2:} Superlevel filtration of the Euclidean Distance Transform (EDT) shown in 2D, 3D, and top views. As the threshold decreases, high-distance regions (deep voids) first appear as isolated components (H0), which grow and merge to form connected structures and eventually loops (H1). The animation highlights the birth, merging, and persistence of topological features.